# Cryogenic Compact mm-Wave Broadband SPST Switch in 22nm FDSOI CMOS for Monolithic Quantum Processors

Tan D. Nhut[1], Shai Bonen[2], Gregory Cooke[2], Thomas Jager[2], Michele Spasaro[1], Dario Sufrà[1], Sorin P. Voinigescu[2], Domenico Zito[1]

[1]Dept. of Electrical and Computer Engineering, Aarhus University, Denmark
[2]Edward S. Rogers Sr. Dept. of Electrical and Computer Engineering, University of Toronto, ON, Canada
domenico.zito@ece.au.dk

*Abstract*—This paper reports the experimental characterization at the cryogenic temperature of a compact mm-wave broadband single-pole single-throw (SPST) switch in 22nm FDSOI CMOS technology. The switch consists of two n-MOSFETs with a special device option to reduce the substrate parasitic effects, and a third n-MOSFET to improve isolation. Unlike prior wideband mm-wave switches, it does not require any large passive components, allowing a very compact design, low loss and high isolation performance. The cryogenic measurements at 2 K show an insertion loss lower than 2.3 dB, an isolation better than 25.3 dB, and the return loss better than -11.5 dB, over the entire frequency range from DC to 70 GHz.

*Keywords*—22nm FDSOI, cryogenic, mm-wave, single-pole single-throw (SPST), switch, wideband.

## I. Introduction

Among the possible candidates for the implementations of large-scaled quantum computers, silicon-based spin qubits have the potential to integrate monolithically the classical control-readout electronics with the qubit devices [1, 2]. Such an integration in commercially available semiconductor technologies is key to move quantum technology out from research laboratories, so enabling very large-scale integrated quantum processors capable to tackle useful applications, which could require thousands and even millions of physical qubits [3]. The monolithic integration of silicon spin qubits together with control and readout circuits, possibly operating at more acceptable cryogenic temperatures, i.e., above 1 K [2, 4, 5], requires a massive number of switches to distribute the control signals to the many qubits and isolate these from the readout circuits during manipulations, as shown in Fig. 1. When targeting scalable qubit control ICs, compact area, wideband operations, high isolation and low insertion loss are essential.

The recent literature reports some wideband mm-wave switches; they make use of inductors/transformers and transmission lines, which allow reaching good performance over frequency [6-14], but with negative impact on silicon area.

In this paper, we report the experimental characterization at cryogenic temperature (CT) of a compact wideband mm-wave single-pole single-throw (SPST) consisting solely of three n-MOSFETs in 22nm FDSOI complementary metal-oxide-semiconductor (CMOS), two of them with a special device option to reduce the substrate effects, a third n-MOSFET to improve isolation, allowing altogether a very compact design, low loss and high isolation performance for effective integration in monolithic quantum processors.

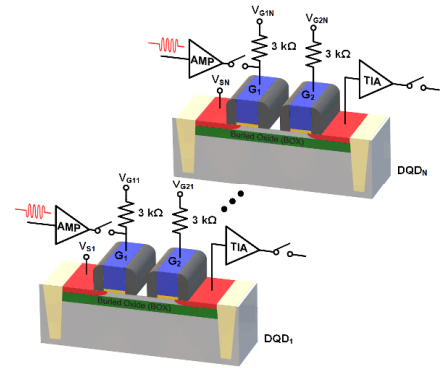

Fig. 1. Example of control and readout of array of N spin qubits based on double quantum dots (DQDs).

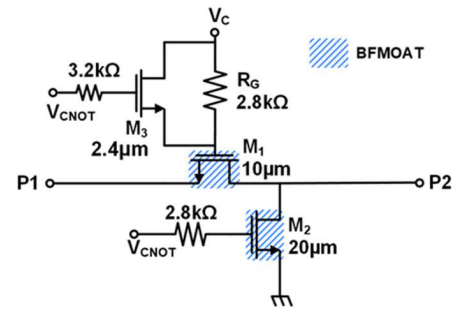

Fig. 2. SPST switch circuit. $M_1$ and $M_2$ are n-MOSFETs with BFMOAT, i.e., a specially treated region for lower losses and higher isolation. The DC control voltages $V_C$ and $V_{CNOT}$ allow switching the device in ON and OFF modes.

The paper is organized as follows. Section II reports the main features of the switch. Section III reports the experimental measurements. Finally, Section IV draws the conclusions.

## II. Switch Features

Fig. 2 shows the compact wideband mm-wave SPST switch based on a series-shunt topology. It consists solely of three n-MOSFETs and does not require spiral inductors/transformers and transmission lines. Such a solution leads to very compact design with reduced parasitics and good wideband performance, as this is not reduced by lossy interconnections (including vias), whose effects are predominant at high frequencies.

The switch has been designed to achieve a low insertion loss (IL), while maintaining an isolation (ISO) of 25 dB at 60 GHz, i.e., the central frequency of the mm-wave frequency band of interest. n-MOSFET devices with a 20nm nominal gate length were used for the implementation.

The commercially available 22nm FDSOI CMOS process includes super-low-$V_T$ (slvt) n-MOSFETs with low channel resistance, leading to lower equivalent resistance between the two poles ($P_1$, $P_2$) and then low IL. For the wideband switch, the roll-off of the IL over frequency is degraded by the intrinsic capacitances of the transistor and extrinsic capacitances of the interconnections. The same technology process provides n-MOSFET device option with a specially treated region (BFMOAT) in order to reduce the parasitic effects in the substrate network [9, 10]. Therefore, the BFMOAT n-MOSFET can be a valid option to reduce the IL roll-off in mm-wave switches.

Fig. 3 shows the cross-sectional view of the BFMOAT n-MOSFET. A thick specially-treated region is implemented below the BOX layer to further reduce the source/drain to substrate capacitances. As the capacitance is proportional to the dielectric constant of the BFMOAT region, we expect that such a capacitance reduction effect holds at CT. Post-layout simulation (PLS) of the BFMOAT n-MOSFET, including five metallization layers, shows an intrinsic switch cut-off frequency $f_C = 1/(2\pi R_{on} C_{off})$ of 1.12 THz at room temperature.

The additional n-MOSFET $M_3$ is introduced in parallel with $R_G$ in order to further improve the isolation when the switch is in the OFF mode. The small channel resistance of $M_3$ bypasses the signal to ground and so improving the isolation. The PLS results showed a 3dB improvement in isolation at 60 GHz, with respect to the case without $M_3$.

The gate width of the BFMOAT n-MOSFETs $M_1$ and $M_2$ have been sized to 10 μm and 20 μm, respectively. The width of slvt n-MOSFET $M_3$ is 2.4 μm. The three-dimensional (3-D) view of the switch layout is illustrated in Fig. 4. The chip microphotograph is reported in Fig. 5.

## III. EXPERIMENTAL CHARACTERIZATION

The switch has been measured on die at room and cryogenic temperatures in order to validate the design carried out at 300 K due to the limited availability and accuracy of models at 2 K.

### A. Room Temperature

The switch has been measured at room temperature with the Keysight Microwave Network Analyzer PNA-X N5245A and characterized over three frequency bands: DC-to-50 GHz, V-band (50-75 GHz), and W-band (75-110 GHz). Stand-alone GSG RF pad test structures were fabricated both for calibration and de-embedding.

Fig. 6 shows the measurement results. In the ON mode ($V_C$ = 0.8 V; $V_{CNOT}$ = 0 V), the switch exhibits an $S_{21}$ of -1.6 dB at 1 GHz and decreases gradually to -2.65 dB at 60 GHz, and -3.1 dB at 110 GHz. The switch exhibits $S_{11}$ lower than -12.1 dB, i.e., the value at 110 GHz, all over the DC-to-110 GHz frequency range. In the OFF mode ($V_C$ = 0 V; $V_{CNOT}$ = 0.8 V), $S_{21}$ amounts to -60 dB, -26.4 dB and -22.8 dB at 1 GHz, 60 GHz, and 110 GHz, respectively. The input-referred 1dB compression point (i$P_{1dB}$) amounts to about -1.8 dBm and 9.9 dBm for input tones of 0.5 and 35 GHz, respectively.

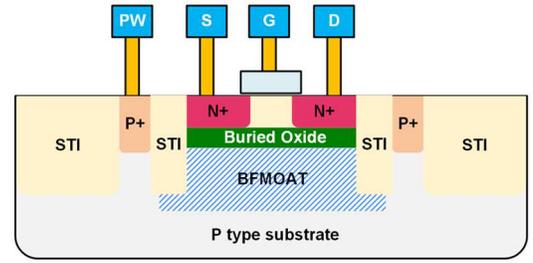

Fig. 3. Cross-section of n-MOSFET with specially treated region [6].

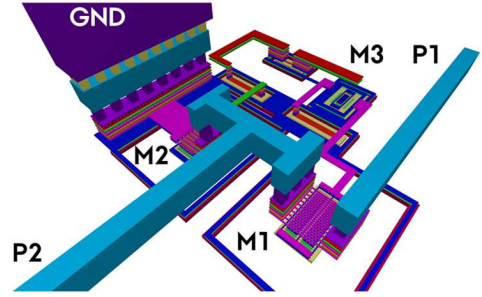

Fig. 4. Illustration of the layout in 3-D view.

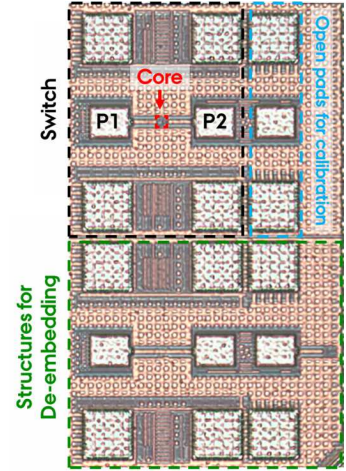

Fig. 5. Chip photograph of SPST switch together with test structures for calibration and de-embedding. The core area amounts to 12.6 × 13 μm².

### B. Cryogenic Temperature

Fig. 7 shows the cryogenic measurement setup from DC to 70 GHz with a Lake Shore Cryotronics CPX-VF-LT probestation. Fig. 8 shows the measurement results. In the ON mode, the switch exhibits $S_{21}$ of -1.3 dB at 1 GHz, and decreases gradually to -2.3 dB at 70 GHz. The switch exhibits $S_{11}$ lower than -11.5 dB, all over the DC-to-70 GHz frequency range. In the OFF mode, $S_{21}$ amounts to -62 dB and -25.3 dB at 1 GHz and 70 GHz, respectively.

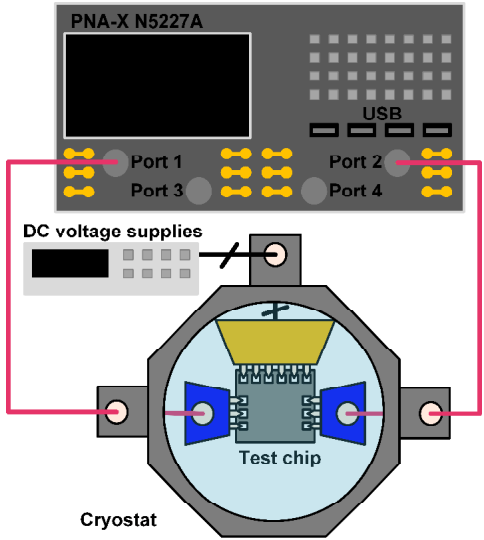

Fig. 7. Closed-cycle helium cryogenic on-chip measurement setup from DC to 70 GHz. These include multi-contact DC and GSG RF probes.

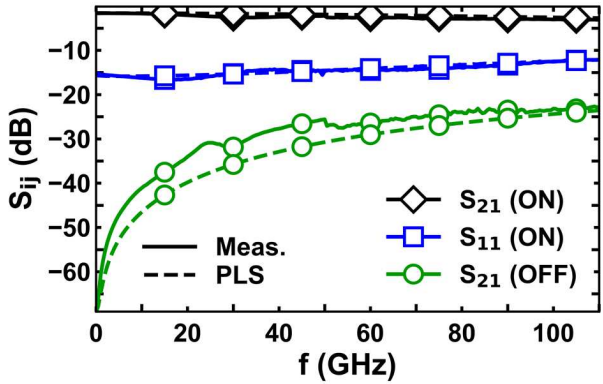

Fig. 6. Measured versus simulated S-parameters of the SPST switch at 300 K.

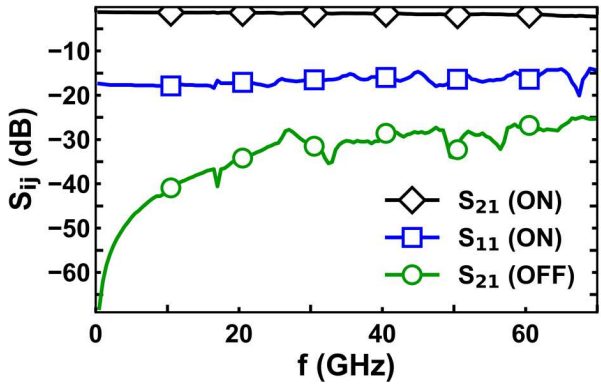

Fig. 8. Measured S-parameters of the SPST switch at 2 K.

Table I. Performance and comparison with prior-art CMOS mm-wave switches.

| Ref. | This work | | [6] | [8] | [9] | [14] |
|---|---|---|---|---|---|---|
| Type | SPST | | SPST | SPST | SPST | SPDT |
| Tech. | CMOS[a] | | CMOS[b] | CMOS[a] | CMOS[a] | SOI-CMOS |
| Freq (GHz) | DC-110 | DC-70 | DC-220 | 10-110 | DC-220 | DC-40 |
| IL (dB) | <3.1 | <2.3 | ≤1.5 | <1.8 | <3.1 | <1.25 |
| ISO (dB) | >22.8 | >25.3 | >16 | >23 | >37 | >20 |
| RL (dB) | >12 | >11.5 | >10 | >10 | >12 | >10.5 |
| $P_{1dB}$ (dBm) | 9.9 | -- | 7 | 7 | - | 12.6 |
| @ GHz | 35 | | 50 | 24 | | 8 |
| Temp. (K) | 300 | 2 | 300 | 300 | 300 | 78 |
| Area[c] (mm²) | 0.00016 | | 0.16 | 0.018 | 0.026 | - |

[a] 22nm FDSOI  [b] 45nm FDSOI  [c] Core

Table I summarizes the performances of the switch at room temperature (300 K) and cryogenic temperature (2 K), together with the performances of prior-art broadband mm-wave CMOS switches. This is the first switch reported at deep cryogenic temperature; it has a record compact area; it exhibits a low IL and high ISO and return loss (RL) comparable with the prior works. The results show that the switch with the BFMOAT n-MOSFETs exhibits good performances also at 2 K, as effect of the reduced capacitance of the BFMOAT region and the increased resistivity caused by the substrate freeze-out.

IV. CONCLUSIONS

A compact broadband mm-wave SPST switch has been designed in 22nm FDSOI CMOS and characterized experimentally at 300 K and 2 K. The switch consists of three n-MOSFETs and does not require any passive components prone to large parasitic effects, in particular, losses, and large area on silicon. Two n-MOSFETs are fabricated on a specially treated region, which allows reducing the substrate parasitic effects; an additional n-MOSFET allows improving the isolation. The cryogenic measurements validate the design and measurements carried out at room temperature, showing good performances. The switch occupies a very small area, i.e., two or more orders of magnitude smaller than prior wideband mm-wave switches, which is crucial in operating on integrated qubits in the form of ultra-scaled quantum devices closely integrated in monolithic scalable quantum processors with arrays of thousands or even millions of physical qubits.

ACKNOWLEDGMENTS

The authors are grateful to Keysight Technologies for their support through the donation of equipment and cad tools; Dr. C. Kretzschmar, Dr. P. Lengo, Dr. B. Chen (GlobalFoundries) for the technology support. This work was co-funded by the European Commission through the European H2020 FET Open project IQubits (G.A. N. 829005).

With reference to Fig. 1, the isolation of the AMP in off-state sums up to the isolation of the switch, reducing the infidelity contribution of the signal leakage to the qubit during an idle state. In other connectivity scenarios, a superior isolation can be achieved by cascading two switches, so leading, in principle, to an overall isolation up to about 50 dB at 70 GHz, with acceptable area penalty owing to the very compact size.

REFERENCES

[1] C. L. Chen, et al, "A Fully Integrated Cryo-CMOS SoC for Qubit Control in Quantum Computers Capable of State Manipulation, Readout and High-Speed Gate Pulsing of Spin Qubits in Intel 22nm FFL FinFET

Technology," *IEEE Journal of Solid-State Circuits*, vol. 56, no. 11, Nov. 2021.

[2] S. Bonen, et al., "Cryogenic Characterization of 22nm FDSOI CMOS Technology for Quantum Computing ICs", *IEEE Electron Device Letters,* vol. 40, no. 1, Jan. 2019, pp. 127 – 130.

[3] A. G. Fowler, et al., "Surface codes: Towards practical large-scale quantum computation," *Phys. Rev. A*, vol. 86, no. 3, p. 032324, Sept. 2012.

[4] C. H. Yang, et al., "Operation of a silicon quantum processor unit cell above one kelvin," *Nature*, vol. 580, no. 7803, pp. 350–354, Apr. 2020.

[5] L. Petit, et al., "Universal quantum logic in hot silicon qubits," *Nature*, vol. 580, no. 7803, pp. 355-359, Apr. 2020.

[6] M. Uzunkol and G. M. Rebeiz, "140–220 GHz SPST and SPDT Switches in 45 nm CMOS SOI," *IEEE Microwave and Wireless Components Letters*, vol. 22, no. 8, pp. 412-414, Aug. 2012.

[7] J. Hoffman, et al., "Analog Circuit Blocks for 80-GHz Bandwidth Frequency-Interleaved, Linear, Large-Swing Front-Ends," *IEEE Journal of Solid-State Circuits*, vol. 51, no. 9, pp. 1985-1993, Sept. 2016.

[8] R. Ciocoveanu, R. Weigel, A. Hagelauer and V. Issakov, "A Low Insertion-Loss 10–110 GHz Digitally Tunable SPST Switch in 22 nm FD-SOI CMOS," *IEEE BiCMOS and Compound Semiconductor Integrated Circuits and Technology Symposium (BCICTS),* Oct. 2018, pp. 1-4.

[9] L. Wu, H. Y. Hsu and S. P. Voinigescu, "A DC to 220-GHz High-Isolation SPST Switch in 22-nm FDSOI CMOS," *IEEE Microwave and Wireless Components Letters*, vol. 31, no. 6, pp. 775-778, June 2021.

[10] S. Yadav, et al., "Demonstration and Modelling of Excellent RF Switch Performance of 22nm FD-SOI Technology for Millimeter-Wave Applications," *European Solid-State Device Research Conference (ESSDERC)*, Sept. 2019, pp. 170-173.

[11] M. Rack, L. Nyssens, S. Wane, D. Bajon and J. -P. Raskin, "DC-40 GHz SPDTs in 22 nm FD-SOI and Back-Gate Impact Study," *IEEE Radio Frequency Integrated Circuits Symposium (RFIC)*, Aug. 2020, pp. 67-70.

[12] C. L. Chen, et al., "Fully Depleted SOI RF Switch with Dynamic Biasing," *IEEE Radio Frequency Integrated Circuits (RFIC) Symposium*, Jun. 2007, pp. 175-178.

[13] C. D. Cheon, et al., "A New Wideband, Low Insertion Loss, High Linearity SiGe RF Switch," *IEEE Microwave and Wireless Components Letters*, vol. 30, no. 10, pp. 985-988, Oct. 2020.

[14] A. S. Cardoso, et al., "On the cryogenic performance of ultra-low-loss, wideband SPDT RF switches designed in a 180 nm SOI-CMOS technology", *SOI-3D-Subthreshold Microelectronics Technology Unified Conference (S3S)*, Oct. 2014, pp. 1-2.